# Deployment characterization of a floatable tidal energy converter on a tidal channel, Ria Formosa, Portugal

Pacheco, A.[1], Gorbeña, E.[1], Plomaritis, T.A.[1], Garel, E.[1], Gonçalves, J.M.S.[2], Bentes, L.[2], Monteiro, P.[2], Afonso, C.M.L.[2], Oliveira, F.[2], Soares, C.[3], Zabel, F.[3], Sequeira, C.[1]

[1]CIMA, University of Algarve Ed7, Campus de Gambelas 8005-139 Faro, Portugal
[2]Centro de Ciências do Mar (CCMAR), University of Algarve, Campus de Gambelas, 8005-139 Faro, Portugal
[3]MarSensing Lda., Centro Empresarial Gambelas Campus de Gambelas Pavilhão B1, 8005-139 Faro, Portugal

**Abstract**

This paper presents the results of a pilot experiment with an existing tidal energy converter (TEC), Evopod 1 kW floatable prototype, in a real test case scenario (Faro Channel, Ria Formosa, Portugal). A baseline marine geophysical, hydrodynamic and ecological study based on the experience collected on the test site is presented. The collected data was used to validate a hydro-morphodynamic model, allowing the selection of the installation area based on both operational and environmental constraints. Operational results related to the description of power generation capacity, energy capture area and proportion of energy flux are presented and discussed, including the failures occurring during the experimental setup. The data is now available to the scientific community and to TEC industry developers, enhancing the operational knowledge of TEC technology concerning efficiency, environmental effects, and interactions (i.e. device/environment). The results can be used by developers on the licensing process, on overcoming the commercial deployment barriers, on offering extra assurance and confidence to investors, who traditionally have seen environmental concerns as a barrier, and on providing the foundations whereupon similar deployment areas can be considered around the world for marine tidal energy extraction.

## 1. Introduction

The hydrokinetic energy that can be extracted from tidal currents is one of the most promising new renewable energy technologies [1]. Despite its huge potential, energy extraction from tidal energy converters (TEC) devices is still in its infancy. The prospects for tidal energy converter technologies very much depend on the specific device concept and how those devices can be optimised to efficiently extract energy, minimizing environmental impacts. Science currently has a very poor understanding of both the hydrodynamics and the ecological implications related with the extraction of energy on coastal environments. In few cases where devices have been deployed the data is highly commercially sensitive and thus not in the public domain and available to the scientific community for research development. The deployment of TECs has also been hindered by a lack of understanding of their environmental interactions, both in terms of the device impact on the environment (important for consenting and stakeholder bodies) and environmental impact on the device (fatigue, actual power output, etc.) which is vital to enhance investor confidence and increase financial support from the private sector. The access to freely available, transparently collected monitoring data from real deployments is paramount both for resource assessments and for cataloguing potential impacts of any marine renewable installation.

This paper presents the results from the deployment of a small-scale tidal current turbine (Evopod E1) in a shallow-water estuarine environment, Ria Formosa Portugal, under SCORE project **S**ustainability of using Ria Formosa **C**urrents **O**n **R**enewable **E**nergy production. This 1:10th scale prototype operated from June to November 2017. The general objective of SCORE is to construct an operational envelope, which can be used by technology developers for design concepts of efficient TECs based on environmental and sustainability principles, contributing to the growth of the blue economy. The deployment site and prototype characteristics are presented in sections 2; section 3 presents the challenges on installing, operating and decommissioning E1 prototype, along with the data collected under the monitoring program; section 4 presents the results obtained, which are fully open access and available for download, following the European Marine Energy Centre (EMEC) standards; and section 5 draws the final remarks and describes ongoing work.

## 2. Experimental settings
### 2.1. Deployment site

The experience with the TEC prototype was performed at Faro-Olhão Inlet, the main inlet of Ria Formosa system (hereafter RF), a coastal lagoon located in the South of Portugal (Figure 1). The RF is a multi-inlet barrier system comprising five islands, two peninsulas separated by six tidal inlets, salt marshes, sand flats and a complex network of tidal channels. The tides in the area are semi-diurnal with typical average astronomical ranges of 2.8 m for spring tides and 1.3 m for neap tides. A maximum tidal range of 3.5 m is reached during equinoctial tides, possibly rising over 3.8 m during storm surges. The wind is on average moderate (3 ms 1) and predominantly from the west. Variance analysis of both tidal and nontidal signals has shown that the meteorological and long-term water-level variability explains less than 1% of the total recorded variance [2]. The lagoon is generally well mixed vertically with no evidence of persistent haline or thermal stratification, which relates to the reduced freshwater input and elevated tidal exchanges i.e. the lagoon is basically euryhaline with salinity values close to those observed in adjacent coastal waters [3].

The deployment site was selected nearby the navigation channel of Faro-Olhão Inlet, Faro Channel, the largest and most hydraulically efficient channel of RF. The depths of the channel in the deployment area range between 4 and 15 m (below msl). Faro-Olhão Inlet is the main inlet of the system, trapping 60% of the total spring-tidal prism of the RF system [4]. The inlet is characterised by strong currents (depth average velocities over 2 ms$^{-1}$ at the inlet throat), especially during ebb. A large difference between flood and ebb duration occurs during spring-tides i.e. ebb duration is shorter and mean velocities are higher. This difference becomes smaller during neap-tides. Due to the narrow inlet mouth (Figure 1A) and the strong tidal current, limited offshore wave energy is reaching the lagoon. Nevertheless, the mooring location could experience fetch dependant waves generated by wind blowing over the lagoon water from the NW or NE directions.

Energy from tides was harvested before at Ria Formosa with tidal mills (XII century) and recent tidal energy assessments determined a mean and maximum potential extractable power of 0.4 kWm$^{-2}$ and 5.7 kWm$^{-2}$, respectively [5]. The RF has attracted research interest in all environmental aspects and hence there is a lot of background literature available about biology, morphodynamics and hydrodynamics. The system is particularly adequate for testing floatable TEC prototypes, and representative of the vast majority of transitional systems where these devices can be used to extract energy to power small local communities.

**2.2.    E1 Evopod 1 kW prototype**

Evopod™ is a device for generating electricity from coastal tidal streams, tidal estuaries, rivers and oceanic sites with strong currents (Figure 2). It is a unique floating solution drawing upon proven technologies used in the offshore oil/gas and marine industries [6]. The 1:10$^{th}$ scale Evopod (E1 hereafter) consists of a positively buoyant horizontal cylindrical body of 2 m length and 0.4 m diameter to which are attached three stabilising fins set in a triangle, tethered to a subsurface buoy. Each fin is approximately 1.2 m height, 0.4 m wide and 0.1 thick. The main body and fins are constructed of steel. When deployed, approximately 0.4 m of the fins are above the water surface. The semi-submerged nacelle has surface piercing struts providing sufficient reserve of buoyancy to resist to the vertical component of the drag force produced by the moorings. The surface piercing struts have a small water-plane area so that the motions of E1 in waves are minimised and do not adversely affect the turbine performance.

A four-bladed 1.5 m diameter turbine made of composite material is attached at the rear of the body and is designed to rotate between 20 and 55 rpm. This result on a maximum blade tip speed of 4.3 ms$^{-1}$, driving a 1 kW permanent magnet AC generator at a rated flow velocity of 1.7 ms$^{-1}$. E1 has a cut-in velocity of 0.7 ms$^{-1}$ and it cannot withstand steady flow velocities larger than 1.75 ms$^{-1}$. The width of each blade is approximately 0.1 m and the depth between the sea surface and the highest point of the rotor is 0.45 m. Hence, the E1 device consists of a fixed pitch 4-bladed turbine driving through a step-up planetary gearbox to a 3-phase multipole permanent magnet generator. The power from the generator feeds a navigation beacon plus an extensive suite of instrumentation measuring the flow speed, voltage, current, torque, revs, temperature, resistor settings, yaw angle and mooring tension. Data records are logged internally and transmitted to a remote PC through GSM communication. Table 1 summarises Evopod™ key discriminators at different scales.

**3.   Methodology**

**3.1. Deployment, operation and decommissioning**

The E1 deployment took place on 8$^{th}$ June 2017. Authorization for deployment was obtained from local maritime authorities following a fast and simple administrative procedure. The device was tethered to the seabed using a four-line catenary spread mooring system (Figure 3A). The flow speeds, wave and wind characteristics at the deployment site were used for the design of the mooring system (Table 2). The moorings consists of chain and galvanised wire mooring lines attached to 4 concrete anchors weighting approximately 1 ton each (Figure 3C). The device is a simple fixed pitch downstream turbine, which aligns freely with the predominant current direction. A load cell was placed for the two south and north lines (Figure 4A), respectively, measuring the tension while E1 is extracting energy. Since the prototype has been deployed for three months, it was not connected to the grid and therefore the excess generated power was dissipated as heat into the sea.

The prototype was installed in collaboration with a local marine services company, which was subcontracted to provide a barge boat equipped with a winch (Figure 4B), essential to lower the anchoring weights at their exact planned location, using RTK-DGPS positioning. The operation was performed at slack tide and involved a staff of ten people, including skippers, researchers, divers and technical operators, supervised by the maritime authorities. The prototype operated at site (Figure 4C) until the 21$^{st}$ November, when it was towed back to the harbour and removed from the water. All the anchoring system was removed except the anchoring weights that remained on site.

### 3.2. Data collection
### 3.2.1. Defining deployment location

Prior to the deployment a baseline marine geophysical, hydrodynamic and ecological database for the pilot site was created. Table 3 summarises the data obtained under SCORE project. The first step of data collection was to complement existent LiDAR bathymetric data of RF and refine the depths at the deployment site. For this task, bathymetric data (Figure 5A) were collected using a single beam eco-sounder (Odom Hydrographic System, Inc. with a 200 kHz transducer) synchronized at 1 Hz with a RTK differential GPS (rover unit model, Trimble R6), though a computer interface running hydrographic survey software (Hypack® 2011, Coastal Oceanographics, Inc.). This allows correcting in real time the tidal and surge levels. A side scan sonar (Tritech StarFish 452F, 450 kHZ) survey was performed to evaluate the presence of priority habitats and characterise the bottom of the deployment area in terms of substrate and the texture type. This characterization allowed the detection of rocky and sediment areas that might be present in the area permitting to choose the best sampling technique for habitat characterization on each bottom type detected.

To fully characterise the 3D flow pattern at the deployment location, an ADCP (Nortek AS Signature 1 MHz) was bottom mounted at a mean water depth of 7.7 m (Figure 5B). Current velocities were measured with 0.2 m cell resolution through the water column, measuring 60 s of data every 5 min bursts. One of the main objectives of the current velocity data collection was to set-up, calibrate and validate a vertical averaged hydrodynamic model (Delft3D) of the entire RF [more details on modelling setup are given on 7], to define the extraction potential at the test case site and to confirm that the deployment location satisfies the prototype constrains (Table 2). From Figure 6, it is evident that flow velocities constraints restricted the locations to place the device. The selected area was then target of a more refine characterisation of current velocities to fully characterize the 3D flow patterns at the deployment location during complete tidal cycles. Those measurements were performed with a Sontek ADCP 1.5kHz with bottom tracking (Figure 5B) by mooring the boat at the exact deployment location (red cross, Figure 6). Velocity components were measured along cells of 0.5 m through the water column, by collecting velocity profiles at each 5 s. Based on those measurements, the estimated E1 electrical power outputs, $P_e$, were calculated using Equation 1:

$$P_e = \frac{1}{2} \rho \eta C_p A_T U_{r\_avg}^3 \qquad (1)$$

where, $\rho$ is sea water density (1025 kgm$^{-3}$); $\eta$ represents generator efficiency, gear losses and shaft losses (90%, 5% and 5%, respectively); $C_p$ is the power coefficient (28% [5]); $A_T$ is the rotor swept area; and $U_{r\_avg}$ represents the flow speed averaged through E1' rotor swept area i.e. by integrating the ADCP velocity measurements along the area through which the rotor blades of the turbine spin.

### 3.2.2. Environmental site Characterisation

Several underwater data acquisition methods were tested to identify their viability of use in a high current condition. The techniques employed should give an overview of the priority habitats and communities of species present in the testing area (Figure 5B). Tested sampling methods included: (1) collection of sediment using a "Van Veen" type grab – intended for the quantification and identification of invertebrate species of infauna and also of epibenthic species that are buried in the sand; (2) bottom trawling with a beam-trawl, following the Water Framework Directive's standards [8], which allowed the quantification of epibenthic species (fish and macroinvertebrates) on mobile substrates; (3) underwater visual censuses (UVC) through transects with SCUBA diving, for the identification and quantification of epibenthic fish and invertebrate on mobile substrates; and (4) video transects with Remote Operating

Vehicle (ROV, SEABOTIX L200 equipped with two forward-facing cameras), which was used to identify/quantify epibenthic habitats and species through the analysis of videos collected during each immersion of the underwater vehicle. Moreover, the interactions between marine mammals, marine turtles, seabird and fish with the turbine were evaluated through visual census and the colonization of the mooring system was assessed by visual inspection of 3 fixed quadrats (10x10 cm) in two opposite anchoring weights (3x3 photoquadrats).

The ROV video sledge was used to characterize an impact zone of 60 x 15 m centred in the tidal turbine. The impact zone refers to the area where the device was deployed i.e. the spatial area limited by the four mooring weights and lines connected to the E1. In this area two 40 m transects were carried out, one during the flow tide and another during the ebb tide (Figure 7). The same procedure was carried out in a control zone, located 50 m apart from the turbine' deployment area, at the same depth range, bottom type and similar hydrodynamic conditions. Transects were perform at low speed (< 1 knot) with the navigation being monitored with differential GPS. The study areas (i.e. impact and control) were surveyed in four time intervals: 3 days prior to the deployment (T-3), and 8 days (T+8), 15 days (T+15) and 63 days (T+63) after the turbine had been installed. Wildlife interaction with the turbine were observed using the same schedule.

Video images collected were annotated using COVER software (Customizable Observation Video Image Recorder, v0.7.2) [9]. Every linear meter a still image was used to visually estimate and quantify the percent-cover of the arborescent bryozoan *Bugula neritina* using ImageJ 1.51j8 [10]. Three additional Gopro cameras were also attached to the video sledge, at 70 cm from the bottom, one central facing downward and two in each side at a 45° angle, with the purpose of creating an orthophotomosaic of the seabed. Benthic invertebrates are often selected as indicators of marine monitoring, because of their sessile nature and life strategies, macrobenthos responds moderately rapidly to anthropogenic or natural disturbances [11]. During the pilot study, *B. neritina* had been identified as a structural component of the benthic community, so the cover percentage of this species was defined as a proxy of the potential disturbances affecting the local fauna. *B. neritina* was also chosen as a target species due to their high abundance, sessile nature and easiness of identification from still images.

Prior to the installation of E1, a baseline measurement of noise level was performed in January 2017. The acoustic data was collected with an autonomous hydrophone, the digitalHyd SR-1, installed on a tripod structure (Figure 5C) at a water depth of approximately 11 m, for 13 full days. A similar acquisition procedure was repeated during the device operation for an interval of 19 full days in August 2017. In both occasions, the equipment was set to record 90 s of acoustic data every 10 minutes, over a frequency band from 0 to approximately 24 kHz. The data analysis consists in obtaining estimates of sound pressure levels (SPL) over the entire acquisition interval, mainly based on statistical indicators both for broadband sound pressure level (SPL) and frequency levels. In November 2017, a complementary data recording was carried out during half of a tide cycle from a boat, by displacing the boat from a flow line passing the rotor.

### 3.2.3. Device performance data

The E1 is instrumented to continuously monitor and log various parameters. The parameters captured during the RF deployment were: flow speed (ms$^{-1}$), shaft speed of rotation (RPM), generator output voltage (Volts) and current (Amps), device compass heading (º degree) and mooring tension (kN). The flow speed past the nacelle was measured using an Airmar CS4500 ultrasonic speed sensor. A C100 fluxgate compass from KVH Industries Inc provided compass heading; while mooring tensions, $F_T$, were measured using 0-5kN load cells supplied by Applied Measurements Ltd.

The above analogue data streams were logged using a Squirrel data logger from Grant Instruments. A two-level gear system was installed in order to reduce the shaft speed during high current velocities. The logger has an alarm feature used to control the load on the generator by switching in and out additional resistors. The timing of the gear changes are logged in the system. The base load resistance on the tidal turbine is a battery charger that is used to maintain charge in the on board battery which powers the logger and instrumentation. This tidal turbine battery charging was supplemented by solar panels. The logger set-up allows specifying different sampling rates and logging intervals. At the beginning, the logger was set to record at 1 Hz. After changing the batteries and solar panels, the acquisition rate was changed to 0.1 Hz to the logger' extend power capacity. The only exception was the flow speed sensor, which sampled always at 5 Hz.

The turbine performance data were then read based on the recorded timestamp. First, the time-series were checked for duplicate times and for inconsistences in the recording time step (i.e. from 1 to 10 sec). Common occurring phenomenon could inflict time drifting of the recording parameters at slightly different timestamps. To counter the aforementioned problem, all parameters were interpolated on a common and fixed time step of 10 sec. Second, all recorded parameters were transformed from the measured quantity (Volts) into the correct units using the calibration equations. Finally, the generated time series were smoothed by applying a moving average filter.

The thrust coefficients, $C_T$, were calculated using load cells data using Equation 2:

$$F_T = F_{T,o} + \frac{1}{2}\rho(C_T A_T + C_s A_s)U_{E1}^2 \qquad (2)$$

where, $F_{T,o}$ depicts the tension force measured by load cells at rest (i.e. at 0 ms$^{-1}$); $C_s$ is the drag coefficient of E1 structure (~ 0.15 [6]); $A_s$ is the E1 cross-sectional area (~$1.15A_T$); and $U_{E1}$ represents the flow speed measured by E1' on-board mounted Doppler. Using the load cells data, the $C_T$ for E1 is obtained by fitting a quadratic drag law of the form $y = Ax^2 + b$, where $y = F_T$, $b = F_{T,o}$, $x = U_{E1}$; and $A = \frac{1}{2}\rho(C_T A_T + C_s A_s)$.

### 3.2.4. Wake measurements

Wake downstream of E1 was characterized at different distances downstream of the channels by combining the use of two ADCPs (Nortek Signature 1 MHz and Sontek ADP 1.5 kHz; Figure 3A). The objective of the wake measurements was to construct the velocity field near the E1 in order to detect, if possible, the spatial characteristics of the wake over different tidal stages, currents velocities and rotor velocities. Those measurements were made for complete tidal cycles using two different techniques: (1) continuous boat-mounted transects; and (2) static measurements at fix positions along the flow axis (Figure 3).

On (1), the boat was manoeuvred through pre-defined lines spaced every 5 m from the rotor until 30 m distance. Measurements were performed using a Sontek ADP 1.5 kHz with bottom-tracking and sampling at continuous mode (e.g. sampling a profile every 5 s). The boat speed was set to the minimum possible in order to assure the best possible data density; but high enough to sample the full area in less than 10 min to assure stationary flow conditions (i.e. constant tidal current). Each set of data was measured at 30 min in order to characterise the spatial and temporal distribution of the wake during the peak flood. This resulted in a total number of 9 timestamps during the 3 hours period around the peak flood. A constant vertical grid was created from 1.1 m depth (i.e. first measured valid cell) up to the maximum water depth with a 0.3 m resolution.

On (2), the boat was used as a floatable platform i.e. the Nortek Signature 1 MHz was operated from the boat by displacing the boat from a flow line passing the rotor (i.e. rotor tail and/or wake centre), collecting measurements every 5 s during 5 min bursts. With an ADCP draft of 0.1 m, a blanking distance of 0.1 m and a cell size of 0.2 m, the first reading is at a 0.3 m depth and the last at 4.5 m depth. This set-up allowed characterizing the vertical profile of E1's wake, which its centreline is at an approximate distance of 1.5 m from the free surface (i.e. approximately the rotor centre). A total of 8 complete sets of wake profiles were measured.

## 4. Results

### 4.1. Deployment location

A bathymetric map of the entire Ria Formosa has been built and is provided on the database (http://w3.ualg.pt/~ampacheco/Score/database.html) as netcdf file using the high resolution LiDAR bathymetry performed on 2011, coupled with bathymetric data from the Faro Port Authority and with 2016' bathymetric surveys performed under the SCORE project. The database also provides the mosaic from the side scan survey where main morphological features can be distinguished (i.e. ripples and mega-ripples). An area of about 6 hectares was surveyed using the side scan sonar technique, revealing a seabed mainly composed of sand, coarse sediments and gravel with a high biogenic component (Figure 8).

Figure 9 shows the vertical profiles of the computed horizontal velocity magnitudes observed for a 14 days interval using the Nortek AS Signature 1 MHz. It is evident the tidal current asymmetry that takes place in the Faro-Olhão Inlet, with ebb currents being significantly stronger than flood currents. This result is important since even modest tidal asymmetry can cause large power asymmetry [12]. These velocity measurements allowed validating the Delft3D model and to select the E1 deployment location (Figure 6). The red cross on Figure 6 marks the E1 deployment site which meets the velocity criteria (velocity range between 0.7 and 1.75 ms$^{-1}$) for around 21% of fortnight cycle. Subsequently, velocity measurements were performed during a spring ebb tide at the deployment location with the ADP Sontek 1.5 kHz, with bottom tracking. Figure 10A shows an example of a time-series contour map of the peak ebb currents at the deployment site, permitting to identify the maximum tidal current velocities that E1 could be exposed; while Figure 10B presents an estimation of the predicted power output using Equation 1. Overall, velocity maximums exceeded the threshold value of ~1.75 ms$^{-1}$ at specific cells, reaching up to ~1.96 ms$^{-1}$. However, the limit was not surpassed when those cell velocities were averaged by the rotor diameter at time intervals of 5 s, resulting on a maximum averaged flow velocity of ~1.68 ms$^{-1}$. It can be observed that the rated E1 capacity is almost achieved at peak ebb (~1 kW, Figure 10B), whereas observed power fluctuations are related to turbulence and eddies propagation.

**4.2. Environmental site characterization**

Several limitations were identified in the methods tested. Bottom trawl showed handling limitations in the turbine' deployment area and since this method needs a minimal operation area it would not allow any impact zonation. Bottom trawl is also an extractive technique and therefore not suitable to assess cumulative impacts over time with successive sampling in a small area. Also, as the study area is located in a Natural Park the use of this method was considered the least appropriate. However, the species list provided was the most comprehensive of all methods tested. Thus, surveys were made just for the characterization of the general study area, ensuring the existence of a reference base species list.

Species inventory from UVC accounted for 31 different species. The epibenthic invertebrate and fish communities were composed of typical and frequent organisms in the soft substrates of Ria Formosa, such as: *Octopus vulgaris*, *Bugula neritina*, *Pomatoschistus microps*, *Holothuria arguinensis*, *Alicia mirabilis*, *Sphaerechinus granularis* or *Trachinus draco*. As expected, strong currents made almost impossible the use of UVC through linear transects. However, this was the only method that have detected a seahorse species that is a vulnerable species. Therefore, random transects were done in the specific study area, mainly for species inventory and collection and for ground-truthing ROV data. The strong currents also made extremely difficult to be precise in the location of dredge's samples for the environmental characterization. Later on, during the operational interval, the high hydrodynamism, the rough bottom and the small area to be sampled implied the increase of deployments; given the low efficiency of the technique in such conditions (2 out 3 deployments were rejected/invalid). Furthermore, the analysis of the samples require several taxonomy expertise, which is more time and money consuming.

Using the ROV in high current areas proved to be a difficult operation. To counteract this problem, the ROV was attached to a sledge and towed along the seabed. This would allow a better control while conducting linear transects and provided a stable platform for additional cameras to be attached. The advantages of ROV compared to regular video cameras are mainly related with its dynamic operability namely the possibility of making adjustments in real time (zooming, changing angles and controlling the light intensity) and the main disadvantages are the initial investment in equipment and piloting skills. Due to this preliminary analysis, it was concluded that for the general environmental and impact assessment the use of ROV/video cameras in a sledge was the most appropriate technique because it was the most practical and non-destructive technique available.

A total of 640 images were annotated thoroughly for the presence and quantification of the coverage of the seabed by *B. neritina*. Percent-cover of this species ranged from 0 to 39.7% (mean: 7.8%) taking into account all images analysed. Mean values of seabed cover increased similarly over time in both survey areas (Figure 11), with slight higher mean values taking place in the turbine area. Percent-cover was found to be significantly different across time in the turbine (Kruskal-Wallis ANOVA on Ranks: $H = 169.253$; $P<0.001$) and in the control (Kruskal-Wallis ANOVA on Ranks: $H = 199.645$; $P<0.001$) areas. However, for both turbine and control areas, the percent-cover of *B. neritina* 3 days before the deployment compared to 8 days after the turbine installation were not considered statistically different (Table 4).

The image analysis of the seabed showed an increase in the percent-cover of the bryozoan *B. neritina* during the study period, from early June to early August. Studies suggest that the temporal fluctuations in the abundance of these colonies is correlated with local weather [13]. In Europe [14], and other locations [15, 16], colonies of *B. neritina* are most abundant in months of warmer water temperature. In addition, under natural conditions, colonies tend to be strongly aggregated, and juveniles settle near mature colonies [17]. The results on the abundance of *B. neritina* agree to a moderate extent with the documented natural patterns. Moreover, the increase in the percent-cover of *B. neritina* was identical in both control and turbine areas suggesting that this pattern was not related with the presence of the tidal turbine, but related to environmental factors. Results of the two-month monitoring period showed no evidence of impact on the seabed that could be directly linked to the installation and operation of the turbine.

An acoustic report with estimates of SPL over the entire acquisition period, mainly based on statistical indicators both for broadband SPL and frequency levels, is provided on the database together with time-series of sound pressure levels and frequency, prior to and during the deployment. From a basic frequency analysis over the entire recording time, it was apparent that the site characterized by two distinct periods over 24-hours intervals, where it was evident that periods of reduced boat traffic at night were interchanged with periods of busy boat traffic during the day (Figure 12A). By means of statistical processing over 1-hour periods, an interval of idle regime (reduced boat traffic) and an interval of busy regime (heavy boat traffic) were precisely established.

The discretization of idle and busy regimes allowed to access the contribution of the tidal turbine operation as a noise source. The site of deployment is close to a traffic route leaving or entering the RF system, and therefore idle and busy regimes were expected a priori to occur. Also, the area of deployment is an area subject to the intensification of water velocities through the fortnight tidal cycle. These two factors are prevailing to the variability observed in the noise level. It is clearly observed that the current speed induced a significant increase on the broadband noise level, especially when current speed peaked to maximum values.

Data collected during E1 operation revealed that the device has minimal potential to generate noise and vibration and therefore does not cause disturbance to the environment. Figure 12B shows a time-frequency representation obtained from the complementary data set recorded on 8$^{th}$ of November 2017, at a position of approximately 5 m upwards from E1, when the current speed was peaking at ~0.56 ms$^{-1}$. The result indicates that the turbine was radiating at least two frequencies, 86 and 170 Hz, where the higher frequency might be a harmonic of the lower frequency. The 170 Hz frequency shows an outstanding from neighbourhood frequencies of about 10 dB, and the 86 Hz frequency shows an outstanding of 10 to 12 dB. Another harmonic at about 340 Hz appears to be noticed.

### 4.3. Device performance data

During its operation lifetime, the device had to be pull out of water for maintenance three times due to various failures that are reported in Table 5. Most of the failures occurred with the logging system, which prevented a continuous data recording and were mainly related to the magnitude of flow velocities during neap tides i.e. here was not enough flow for the turbine to generate and feed voltage to the logger. Figure 13 exemplifies the data recorded by the E1 logger over a spring-neap tidal cycle. From top to bottom the following parameters are presents: (i) drag force recorded by the two load cells; (ii) generated voltage (Volts); (iii) generated amperage (amp); (iv) electrical output (Watts); (v) current speed (ms$^{-1}$); (vi) raw power (Watts), i.e. $P = 0.5\rho A_T U_o^3$; and (vii) efficiency in power extraction (i.e. electrical outputs divided by raw output). The shaft speed in rounds per minutes is also logged but the data quality is not the expected, hence data is not presented. In general, and during the peak currents of the spring tides, the shaft speed normally exceed 100 rpm; this values drops to 70 rpm for neap tides. The two load cells values are strongly modulated by the tidal stage. Over spring tides, the drag force can reach to 1 kN. The drag drops to ~0.5 kN during neap tidal ranges. The north mooring under tension during the ebb stage of the flow, which is normally characterized by stronger tidal currents, results in higher load cell values, both in terms of peak values and duration.

Computed thrust coefficients, $C_T$, are illustrated in Figure 14A. Mean computed values of $C_T$ are 0.44 and 0.4 for load cell South and North, respectively. Larger variation of $C_T$ values are observed for flow speeds below 0.5 ms$^{-1}$. When flow speed increases, $C_T$ values converge to mean values. This phenomenon can be explained due to the fact that at higher flow velocities an onset of turbulence in the boundary layer decreases the overall drag of the device. The fitting of a quadratic drag law (Figure 14B) to the measured mooring lines tensions shows that a constant $C_T$ of 0.4 provides a good agreement with the observed flow speeds.

The electrical parameters voltage and amperage, as well as associated electrical output, are strongly related with more than 100 W produced during spring tidal ranges (Figure 13). This production quickly drops within a couple of days from the larger spring tides. For the rest of the tidal cycles the electrical productions are less than 50 W, or even smaller at the neap cycles. As expected, the associated tidal currents speed measured from the E1 Doppler sensor are strongly associated with all the above parameters. In fact, and taking as example the electrical output, it is observed that for velocities less than 1 ms$^{-1}$ the produced power drops by a factor of 2. For the same time, the raw power was of the order of 1 kW over the most productive tide phases, dropping to half when the peak tidal currents did not exceed 1 ms$^{-1}$.

Regarding E1's operating efficiency, the recorded values during the deployment (Figure 13) differ from the power curve provided by the manufacturer and calculated using a constant power coefficient, $C_p$ = 28 %, resulting in a $\eta C_p$ = 22 % (Figure 15A) i.e. although the maximum efficiencies observed are of 23 % at 0.8 ms$^{-1}$, slightly higher than the value of 22 % specified in Equation 1 (i.e. $\eta C_p$), average values are of ~9 %. Overall, efficiencies larger than 15% are observed at flow speeds below 1.1 ms$^{-1}$ (Figure 15B). Above this flow speed, efficiency starts to drop to an average value of ~6%. For the highest flow speed, ~1.42 ms$^{-1}$, efficiency is ~5.4%. These low efficiency values, and the tendency of efficiencies' decrease with increasing flow speeds, can be related to the load control system of the generator and to flow speed fluctuations. When switching in and out the resistors due to variations on flow speeds causes abrupt oscillations on power output affecting the device' efficiency. It is important to remark that the power curve provided by the manufacturer is calculated assuming a constant power coefficient ($C_P$ = 28 %), when usually power coefficients vary with flow speed.

## 4.4. Wake measurements

The 2D wake field was measured with the Sontek 1.5 kHz ADCP operated from moving the boat along the deployment area (Figure 3). Since the rotor centre is about 1.5 m below water lever, the rotor blades spins between approximately 0.75 until 2.25 m depth. The ADCP cells within this range were vertically averaged to compute the wake effect of the E1. However, and because of the restrictions imposed by the sampling rate and boat velocity, the spatial distribution of the ADCP profiles were not optimal for a detail mapping of the wake. The relative strong current velocities make difficult the boat' navigation resulting in a more random distribution of the sampling points. In addition, the flow velocity near the E1 is also characterised by turbulent flow. Those turbulences cannot be spatially and temporarily averaged due to the sampling restrictions mentioned above.

During the peak of the flood currents, some wake patterns can be identified by combining the horizontal and the vertical velocities field, averaged over the vertical layers situated at the blade spinning area (Figure 16). There is evidences of an unsteady pattern on the horizontal components. Although is not a clear wake signature, the vertical component shows an increase of the l module at the expected wake positions, most likely caused by the blade rotation. It is also likely that the presence of horizontal eddies on the ambient flow are masking the wake signal.

Complementary, the static wake measurements along E1's wake centreline obtained with the Nortek AS Signature 1 MHz ADCP for a full profile are presented on Figure 17A. Figure 17B summarises the wake velocity deficits for all measured profiles (i.e. $U/U_o$, relating flow velocities with the presence of the turbine, U, and without turbine $U_o$) at the rotor horizontal plane' height, for each E1' downstream location (i.e. 5 m, 10 m, 15 m, 20 m, 25 m and 30 m). From Figure 17B, it can be seen how the wake re-energizes gradually downstream E1 and recovers almost completely at a distance of 30 m (i.e. 20 rotor diameters). Immediately behind E1, the wake's vertical distance matches the diameter of the turbine rotor (i.e. 1.5 m). The distortion of the velocity profile caused by the wake expands progressively at each downstream distance and the minimum flow velocities are found at deeper depths, until the velocity profile recovers it normal shape. This wake recovery pattern is observed in all measured profiles (Figure 17A). From the box plot (Figure 17B), it is observable that the velocity deficits varied from 0.8 (first quartile) at 5 m to 0.97 (third quartile) at 30 m downstream. Median values of velocity deficits increase with distance as wake recovers. At closer distances downstream E1, it is sensed a larger deviation of velocity deficits. This can be related to the fact that, in the near wake, velocity gradients are larger and its width is shorter than in the far wake. Thus, at these locations, small changes in the lateral position of the ADCP produce larger variabilities on the measured velocities. Wake measurements were only conducted during flood tide, so the wake characterization did not account for any directional asymmetry between the flood and ebb currents.

## 5. Final remarks

Prototype testing of TEC devices is an extremely important part of proving that they will function in full-scale conditions; on the other hand, understanding their potential environmental impacts is a key issue in gaining acceptance of new technologies. Currently little is known about the environmental effects of TEC devices particularly when deployed in semi-closed systems such as coastal lagoons and estuaries. Uncertainties associated with scaling up the impacts from pilot scale to commercial scale are undocumented for floating tethered TEC. The innovative aspect of E1 testing in Portugal laid with the unique morphological characteristics associated with the device deployment site at RF, a coastal lagoon protected by a multi-inlet barrier system. The E1' testing allowed the collection of a significant amount of data (Table 3) that are now available for the science community. The paper also reports the problems (Table 5) occurred during the device testing, essential to wider the understanding of the challenges imposed by extracting energy at these locations and with these equipment. Some key lessons were highlighted:

(1) The existence of data characterising environmental conditions prior to extraction of energy at any location is essential for cataloguing potential impacts of any marine renewable installation [18]. Primary concerns relating to TEC installations are interference with the local ecosystem during installation activities, the potential of the rotating blades to injure fish, diving birds and sea mammals and the loss of amenity i.e. habitat loss due to noise, fishing areas and navigation space for other users of the sea area [19-21]. No collisions or major interactions occurred with wildlife and mooring weight were rapidly colonized by the typical species normally present in the area;

(2) The high energy environment coupled in a restricted work area, heavy chain moorings and a tidal turbine with rotating blades made the use of traditional biological sampling techniques a challenging task. Among the methods tested, the video sledge proved to be the most reliable to be used in these demanding environmental conditions. Complemented with visual census during neap tide, this method was considered the most consistent and replicable technique for the biological characterisation and the following monitoring period while device was operating;

(3) The results from the assessment of the soft sediment community in the study area during the monitoring period did not show signs of disturbance that could be directly linked with the presence of neither the turbine nor the mooring system used. The effect of mooring lines on the seabed is restricted to a few centimetres at both sides of the mooring lines;

(4) The species chosen as a bioindicator, *B. neritina*, despite being considered an invasive species, has a wide distribution in the area of deployment and surrounding area, is a sessile benthic organism and among the fauna present was the most common and conspicuous organism. Their increase in abundance was more related with abiotic conditions during the monitoring period rather than short-term probable impacts caused by the tidal turbine. Future studies should take into account long term monitoring to provide a better overview of the potential impact of this kind of structures. Since no evidence of impact related to the tidal turbine was detected, it is not possible to infer about cumulative impacts caused by a network of these type of structures;

(5) The background noise level was analysed by means of time-frequency representation, and the investigation of the influence of the tide on the background noise was carried out using the flow speed data. The results of the operational noise of the turbine were then compared to the background noise level. During the peak of tidal current, for an interval of approximately 25 min, the turbine radiated a signal with a fundamental harmonic of approximately 86 Hz, where up to three multiples (second to fourth harmonic) could be seen. The first and second harmonics are relatively energetic, with an outstanding of 10 dB above background noise. The amount of acoustic energy introduced into the aquatic environment is limited in frequency band and time. Yet, further analysis is required to conclude on the acoustic impact in the surrounding area and how it would extrapolate if an array of floatable TECs in real-scale were to be deployed;

(6) Floatable devices have advantages on reducing physical environmental impacts. Because they extract energy from the top surface, they cause less impact on both flow and bed properties. Overall, the physical environmental impact from E1 small-scale TEC pilot project was found to be reversible on decommissioning, especially because the chosen area is characterized by a high current flow that already causes natural disturbances to the bed. No record of any change on the bed related with alteration of either flow or sediment transport patterns;

(7) Floatable devices are tethered to the seabed and under direct impact of waves and surface wind, causing a range of different problems and new challenges to successful extract energy. The exact calculation of mooring loads using safety factors was essential to the success of the deployment. However, the miscalculation of the exact location of one of the mooring weights caused over tension on one of the mooring lines, which interfered with the reposition of the device when turning until the tension was corrected by lengthening the mooring line;

(8) The flow field around turbines is extremely complex. Variables such as inflow velocity, turbulence intensity, rotor thrust, support structure and the proximity of the bed and free surface all influence the flow profile. The majority of flow field studies around tidal turbines have been carried out in laboratories [22-24] i.e. in the few cases that devices have been deployed and monitored data are highly commercially sensitive and not distributed to the public and research community [25, 26]. A full characterisation of the 3D flow patterns was performed using ADCPs (moored and boat-mounted surveys). The data collected allowed validating a numerical modelling platform, essential to accurate positioning the device based on the environment/device constraints, mainly in which concerns cut-in/cut-off velocities and deployment depth. The static measurements performed during device operation were effective on characterizing the wake at different distances from the device and represent a valid data set for wake modelling validation;

(9) E1 proved to be easy to disconnect from the moorings and it transport inshore for maintenance and repair was relatively straightforward. This is an important aspect, since installation/maintenance costs represent a major drawback of TEC technologies for future investors. Biofouling can be a major issue affecting performance of devices operating in highly productive ecological regions like RF. Therefore, maintenance operations need to be planned in advance to control the lifespan of antifouling coatings, especially on the leading edge of blades. Another important aspect is to provide on-site access to the power supply batteries, this way there is no need to take the device onshore for maintenance of internal batteries, which translates in reducing equipment downtime and maintenance costs;

(10) Model data is essential for future planning and testing floating TEC prototypes on other locations by providing values of turbine drag, power coefficients and power outputs for different flow conditions and operating settings [27]. Mooring loads and flow speeds data allowed to calculate time-series of E1 drag coefficient. By fitting a quadratic drag law a constant drag coefficient of 0.4 was obtained for flow speeds up to 1.4 ms$^{-1}$. In order to confirm this estimation it will be necessary to measure mooring tension loads at higher flow speeds;

(11) The operational data collected during the operational stage allowed the monitoring of device performance and serve as basis for developing advanced power control algorithms to optimise energy extraction under turbulent flows. The measured energy extraction efficiency and mooring loads of the operational prototype can now be compared against numerical models in order to validate these tools. Time series of measured efficiency revealed an overall underperformance of E1 respect to its power curve estimations with values of $\eta C_P$ below 20% most of the time. Further research has to be conducted to accurately identify the causes of low efficiencies and determine if the problem is related with mechanical, electrical and/or generator losses. A preliminary diagnose points to the generator's resistors control strategy, which needs to be optimised to increase electrical power outputs when operating in turbulent flows;

(12) Efficiency data obtained with E1 prototype can be scale up for proposing realistic tidal array configurations for floating tidal turbines and on supporting the modelling of mooring and power export cabling systems for these arrays. Those validated modelling tools can then be used for performing simulations using different hydrodynamic settings and number of prototype units in different tidal stream environments. By incorporating single devices and multiple array devices on the modelling domain it will enable energy suppliers to gain a realistic evaluation of the supply potential of tidal energy from a specified site. As an example, drag forces measured by the load cells can help on avoiding over engineering and on developing alternative tension-tethered mooring solutions to allow closer spacing of turbines (i.e. reduce project costs and smaller array footprint);

(13) Finally, Ria Formosa is an ideal place for testing floatable TEC prototypes, and can be used as representative of the vast majority of coastal areas where TECs can be used in the future. In particular, the selected test site, Faro Channel, is an attractive case study for implementing TECs because is characterised by strong currents. The channel is also located between two barrier islands and can be easily connected to the national grid system.

## ACKNOWLEDGEMENTS

The paper is a contribution to the SCORE project, funded by the Portuguese Foundation for Science and Technology (FCT – PTDC/AAG-TEC/1710/2014). André Pacheco was supported by the Portuguese Foundation for Science and Technology under the Portuguese Researchers' Programme 2014 entitled "Exploring new concepts for extracting energy from tides" (IF/00286/2014/CP1234). Eduardo G-Gorbeña has received funding for the OpTiCA project from the Marie Skłodowska-Curie Actions of the European Union's H2020-MSCA-IF-EF-RI-2016 / under REA grant agreement n° [748747]. The authors would like to thank to the Portuguese Maritime Authorities and Sofareia SA for their help on the deployment.

**FIGURE 1**

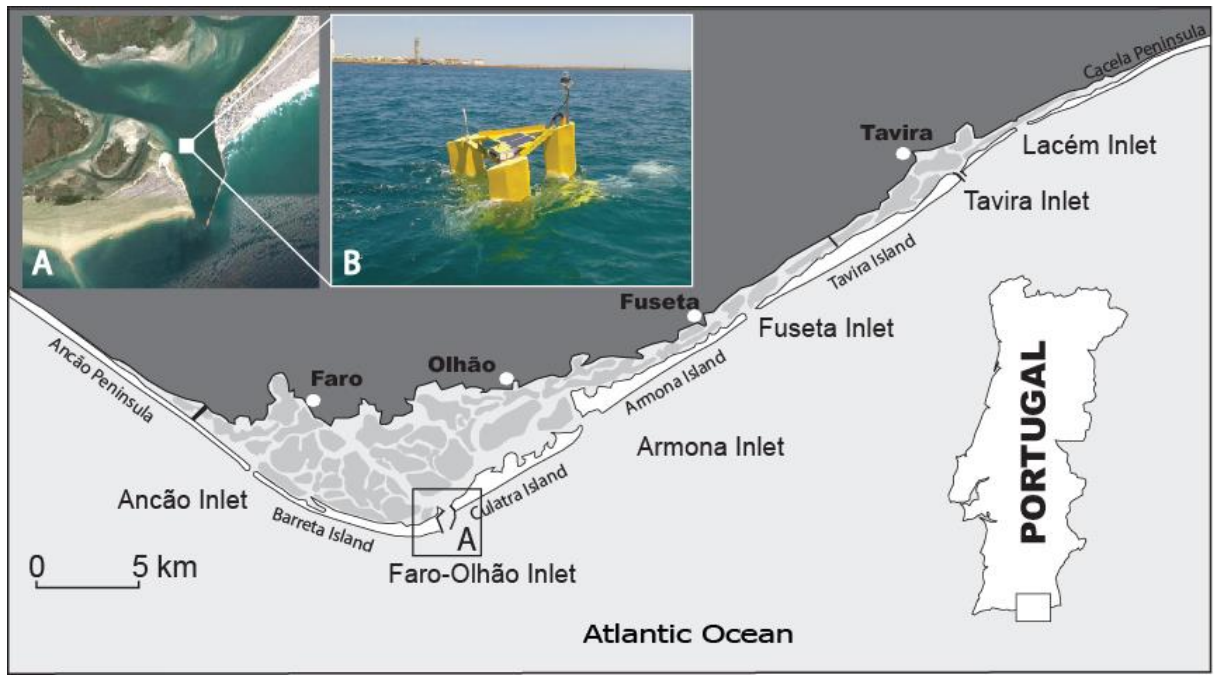

**Figure 1.** Deployment site adjacent to Faro-Olhão Inlet (A), the Faro Channel of Ria Formosa lagoon system (Algarve, Portugal), where E1 Evopod (B) operated. The channel is generally oriented NW–SE, has a length of 9 km, and covers an area of 337 km$^2$. The channel width is not constant, ranging from ~175 m to a maximum of ~625 m. The typical maximum depths along the channel range between 6 and 18 m (below mean sea level).

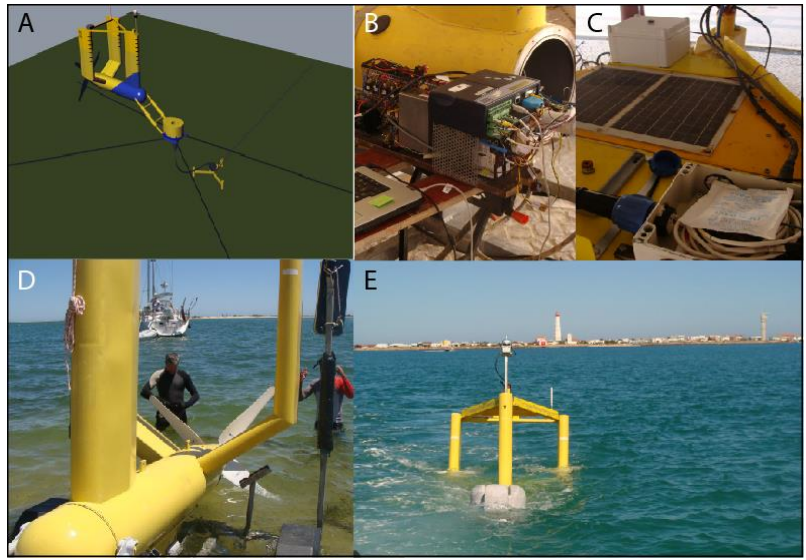

**Figure 2.** (A) Scheme of E1 with the mooring lines spreading from the mid-water buoy; (B) inside components connect to the squirrel logger; (C) detail of the deck with the solar panels and control box; (D) E1 launch on the water and (E) it trawl to the deployment site.

**FIGURE 3**

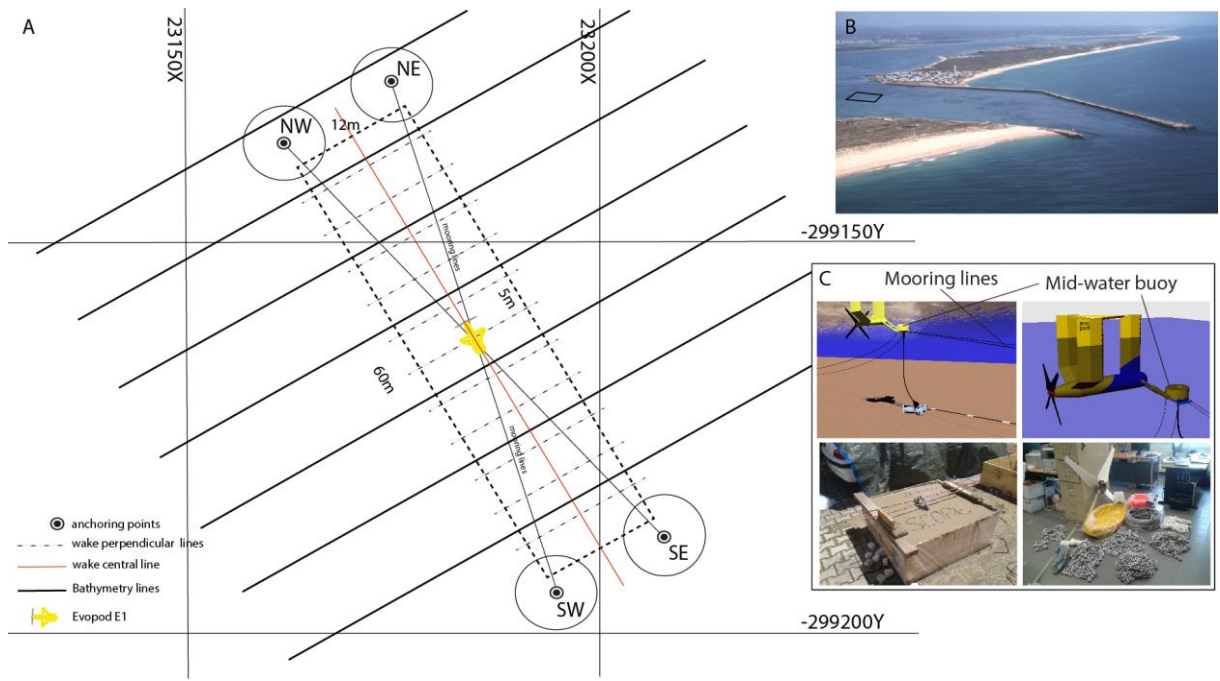

**Figure 3.** (A) Scheme of the deployment site, with the mooring locations and line spreading. Also represented are the bathymetry lines 10 m spaced, the wake perpendicular lines and wake central line where bottom-tracking ADCP and static measurements were performed, respectively; (B) Deployment area represented over an oblique image of Faro-Olhão Inlet; (C) mooring scheme and material used on the deployment (e.g. anchoring weight, chains, marking buoys, cable wire, etc).

**FIGURE 4**

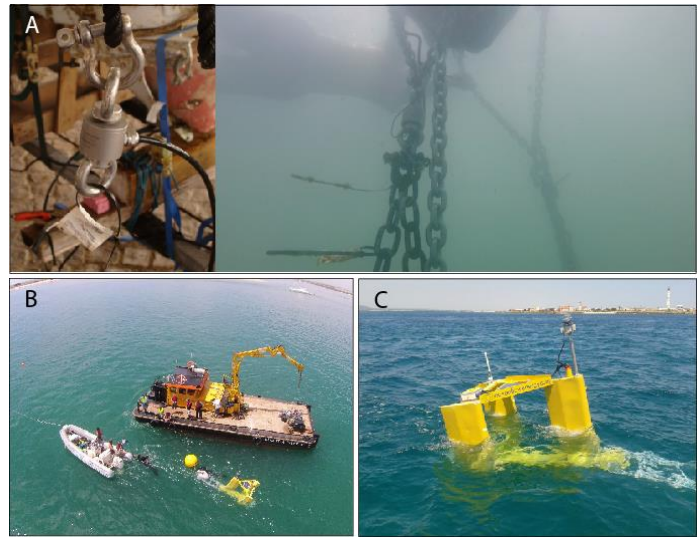

**Figure 4.** (A) Detail of the load cells and its placing on the mooring lines; (B) Deployment day and boats used on the mooring operation; (C) E1 deployed on 8$^{th}$ June 2017.
27

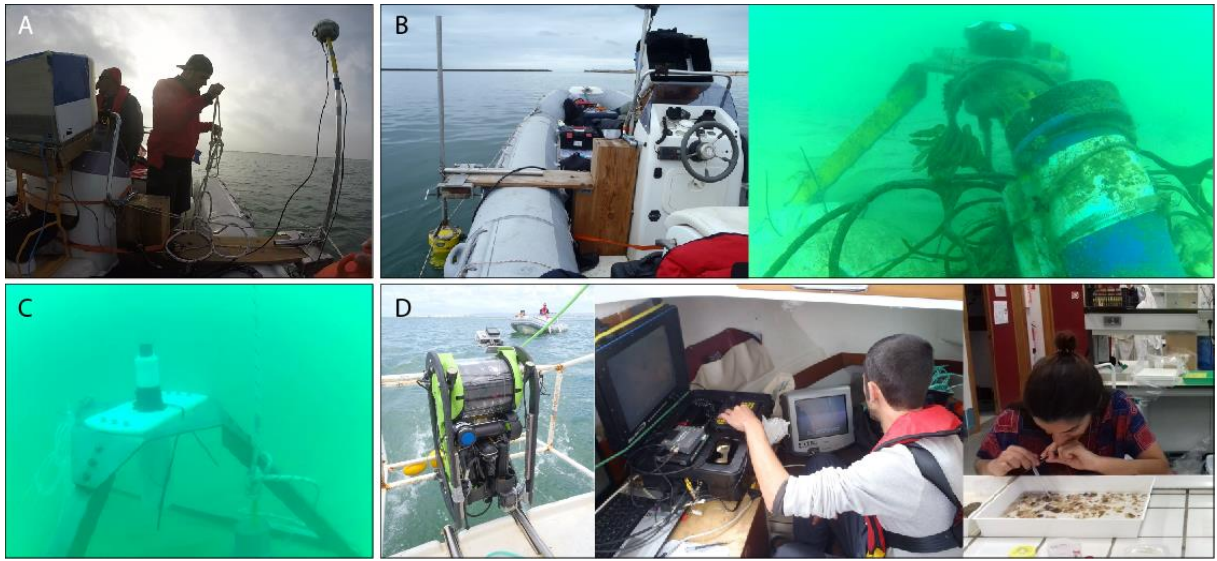

**Figure 5.** (A) Bathymetric survey using a RTK-DGPS synchronized with the single beam echo-sounder; (B) Characterization of the 3D flow pattern using boat mounted (with bottom tracking) and bottom mounted ADCPs; (C) Acoustic measurements with a hydrophone bottom mounted; and (D) ROV videos a for habitat characterization.

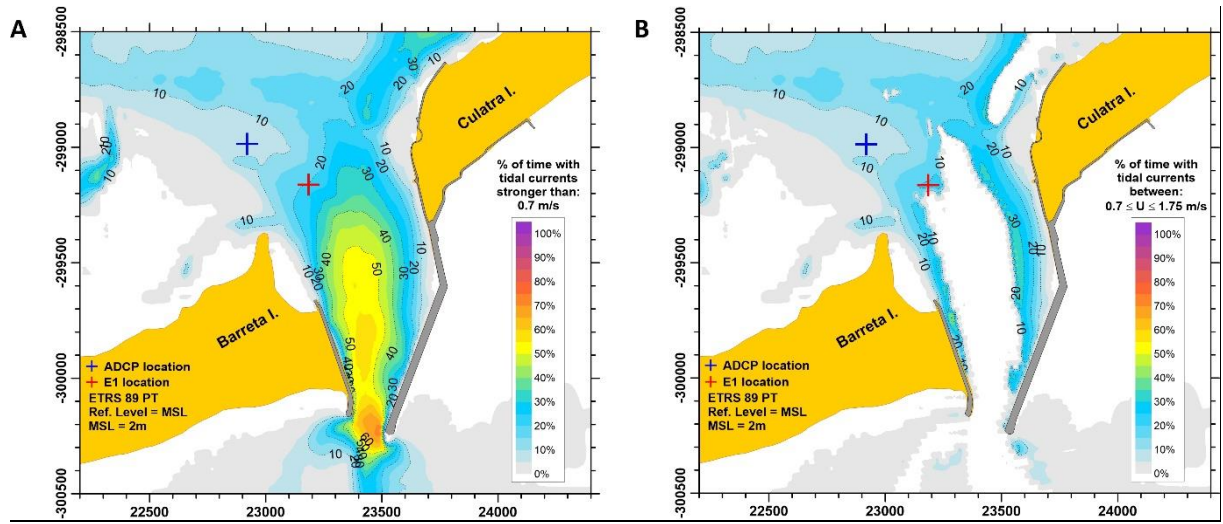

**Figure 6.** Percent of time during a 14 period simulation with occurrence of tidal currents for the Faro-Olhão Inlet area: A) with velocities stronger than 0.7 ms$^{-1}$, and B) with velocities stronger than 0.7 ms$^{-1}$ and lower than 1.75 ms$^{-1}$.

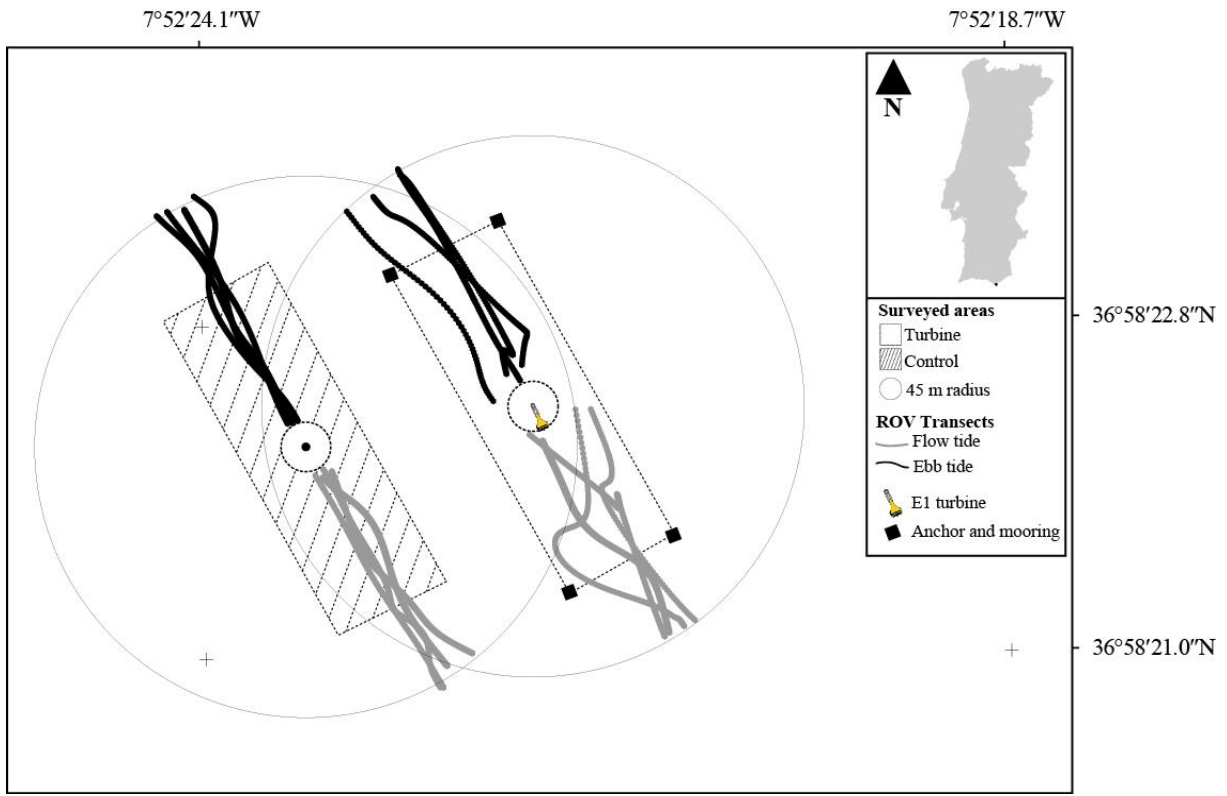

**Figure 7.** Location of the ROV transects carried out in the survey areas during each tidal regime.

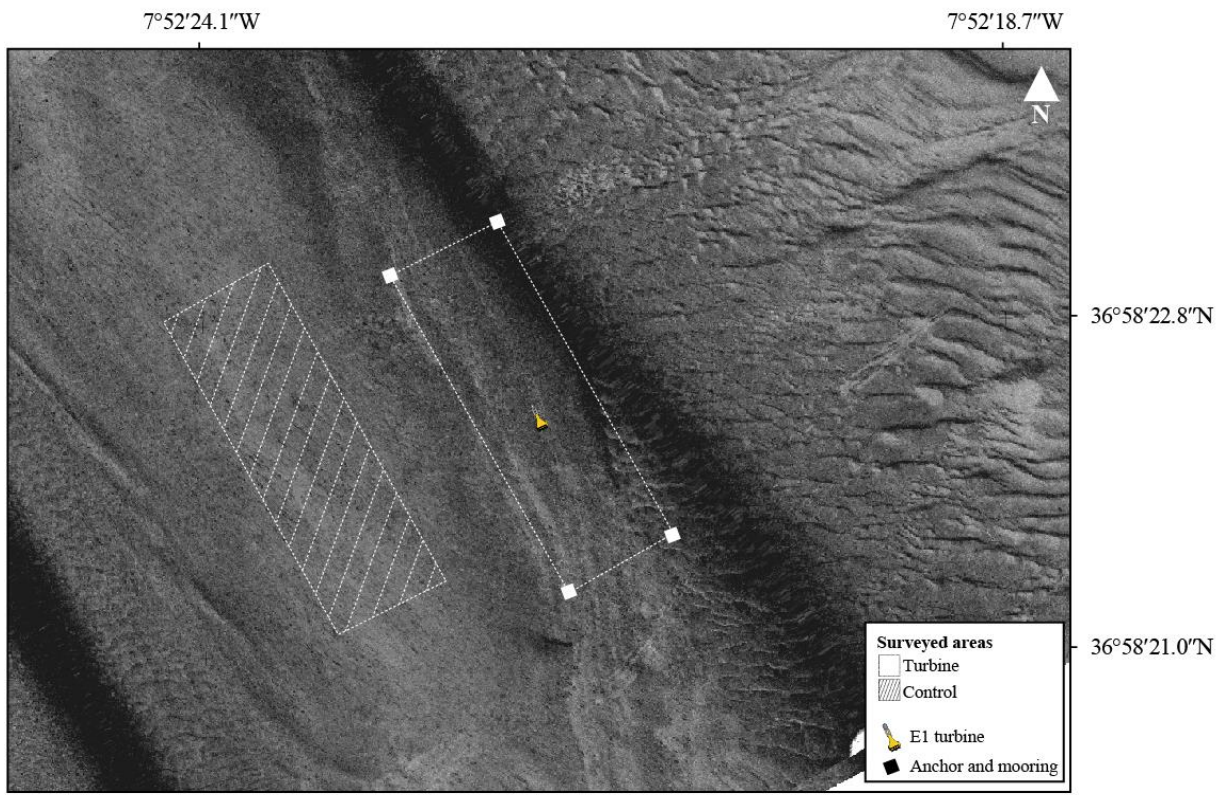

**Figure 8.** Side scan sonar mosaic of the study area.

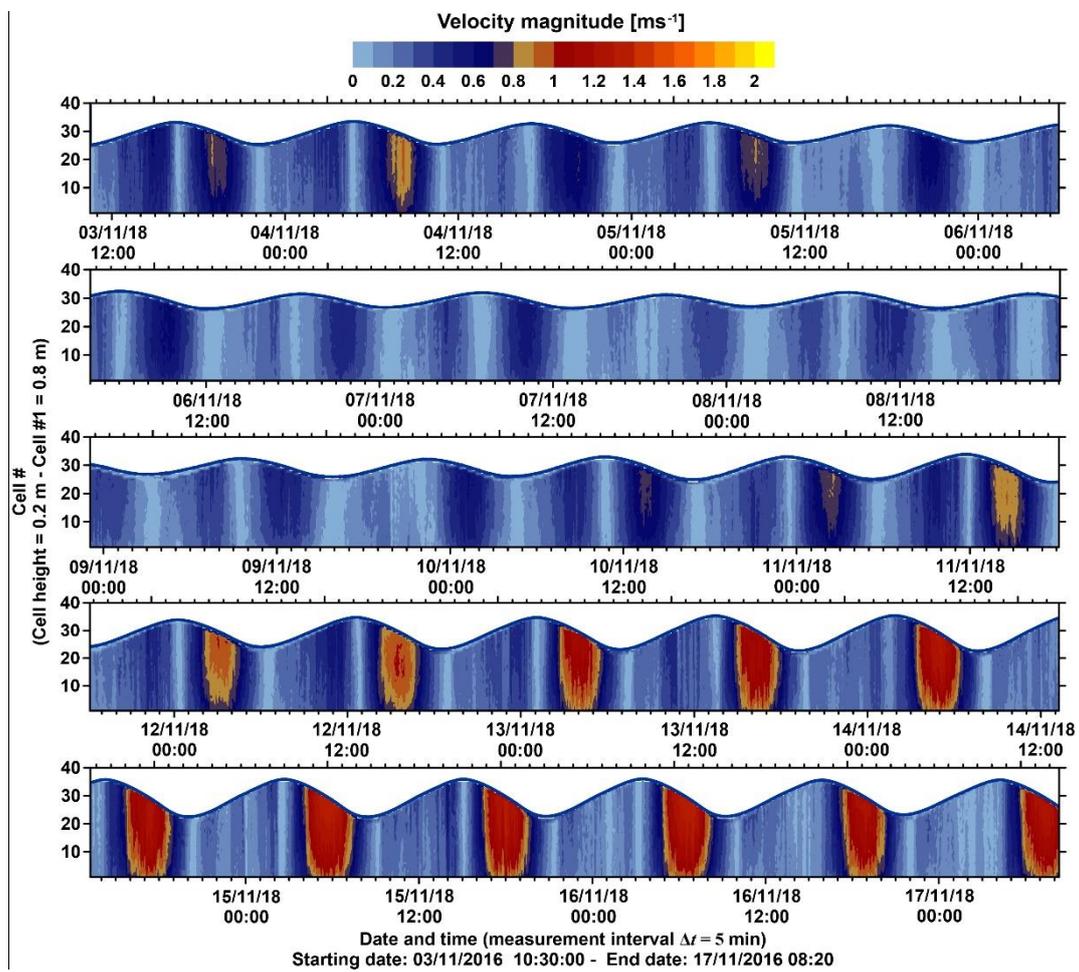

**Figure 9.** Time-series of computed horizontal velocity magnitudes at each cell collected with the Nortek AS Signature 1Mz.

**FIGURE 10**

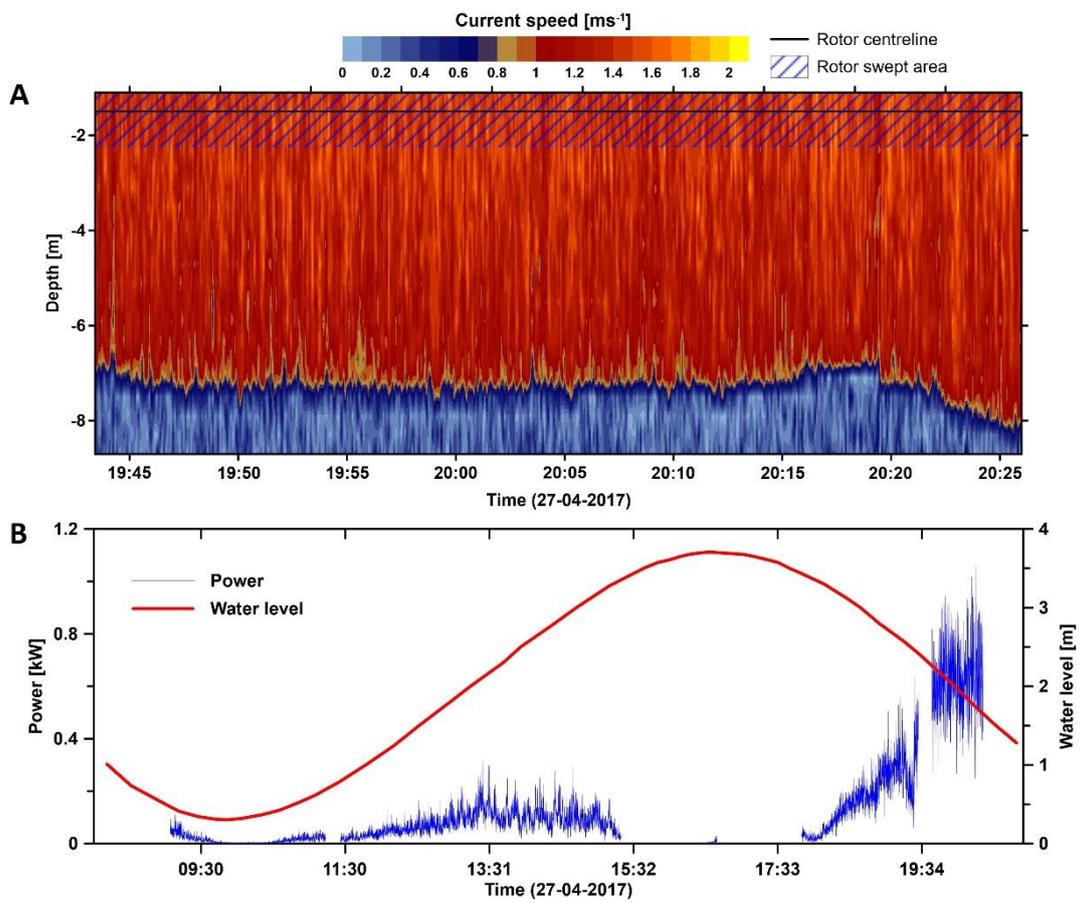

**Figure 10.** (A) Peak ebb current velocities measured at E1 deployment site. Each profile corresponds to an ensemble collected with the Sontek ADCP 1.5kHz with bottom tracking at a 5 s interval; (B) estimated electrical power output for E1 based on the ADCP measurements for a flood-ebb spring-tide (red line).

**FIGURE 11**

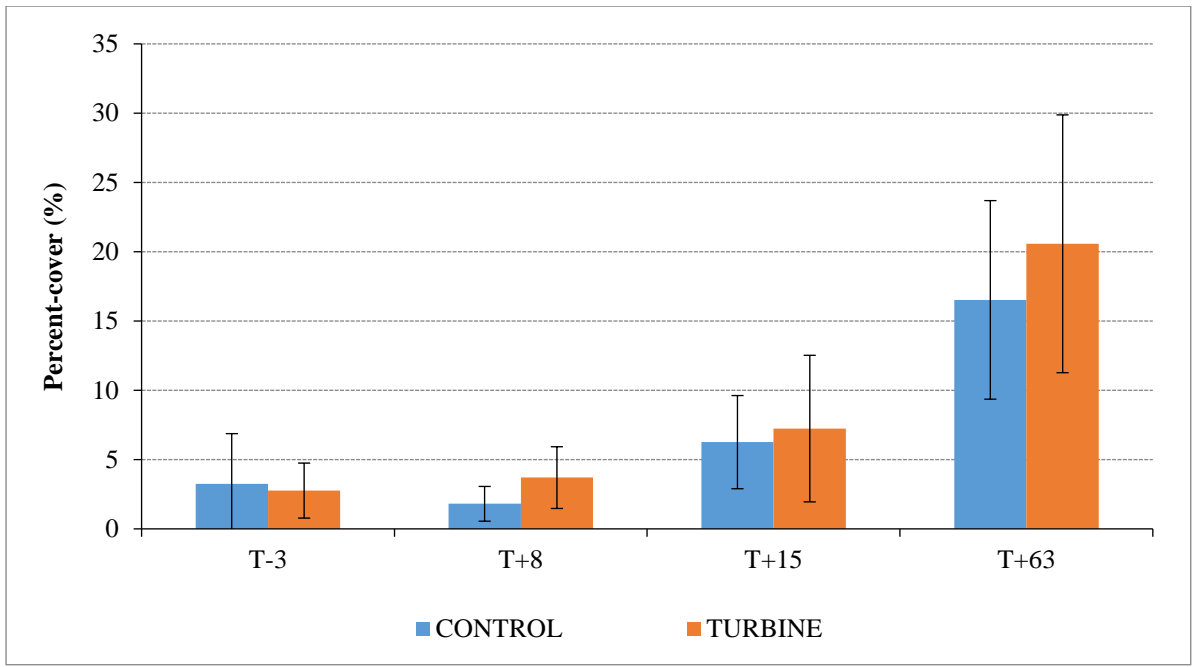

**Figure 11**. Mean values (± standard deviation) of the percent-cover of *Bugula neritina* in both areas surveyed during the study period. T-3: 3 days before to the deployment; T+ 8, T+15 and T+63: 8, 15 and 63 days after the turbine had been installed.

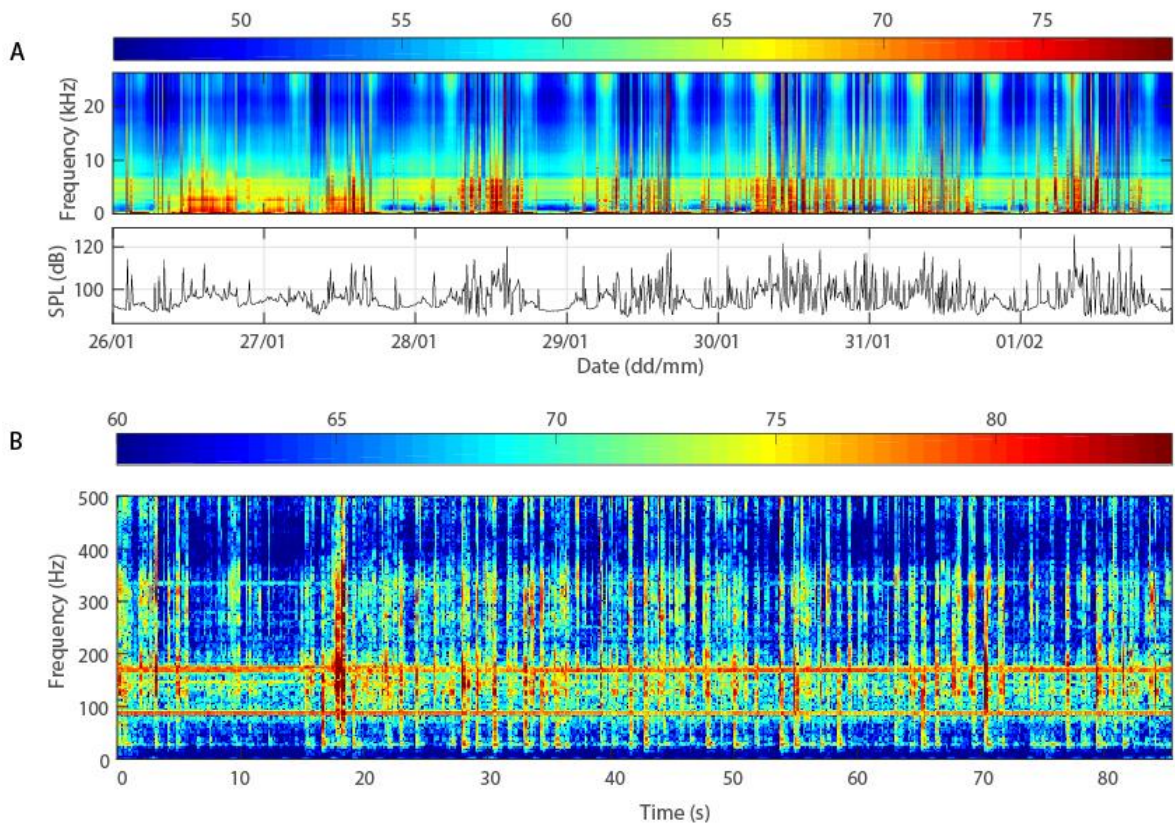

**Figure 12.** (A) Time-frequency analysis of the time series collected from 26[th] January to 1[st] of February 2017 by means of an autonomous hydrophone mounted on a tripod (only the 2[nd] half is shown). The analysis has been performed using observation windows of 4096 samples (≈ 0.077 s) which have been averaged to 90 s using the Welch method; (B) Time-frequency analysis data collected at time 15:12 at 8th November 2017 by means of an autonomous hydrophone operated from a boat. The analysis has been performed using observation windows of 16384 samples (≈ 0.311 s).

**FIGURE 13**

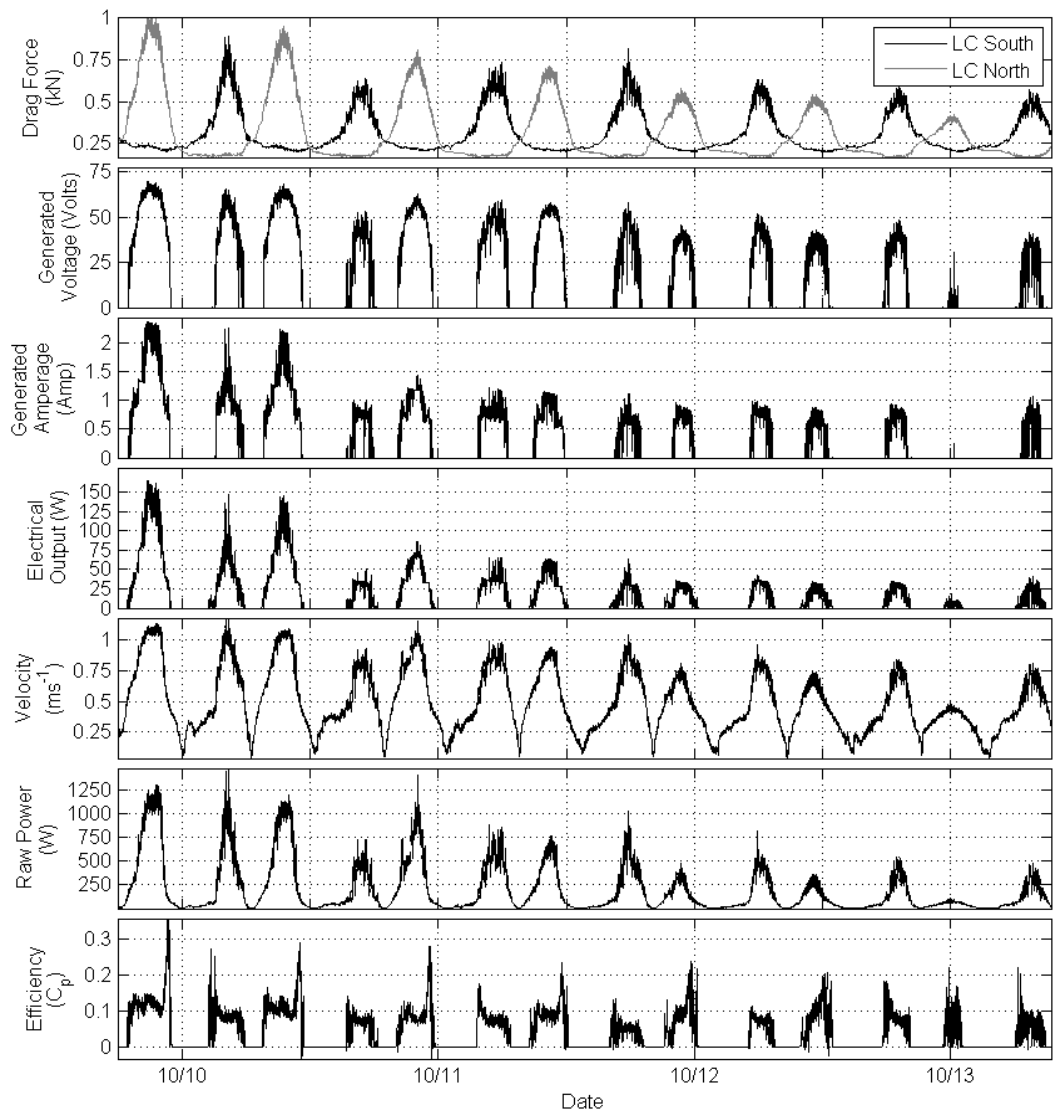

**Figure 13.** Time series of E1 parameters during a spring-neap tidal cycle. From top to bottom the panels present: (i) drag forces recorded by the load cell; (ii) generated voltage; (iii) generated amperage; (iv) electrical output; (v) current speed; (vi) raw power and (vii) E1 efficiency.

**FIGURE 14**

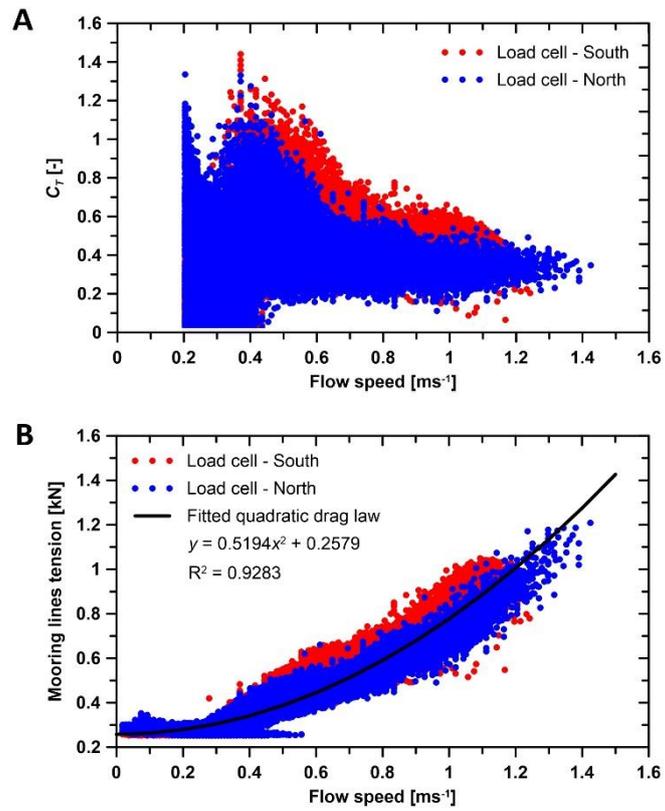

**Figure 14.** (A) Computed $C_T$ for both load cells placed at E1 moorings; (B) observed tension forces for both load cells and fitted quadratic drag law.

**FIGURE 15**

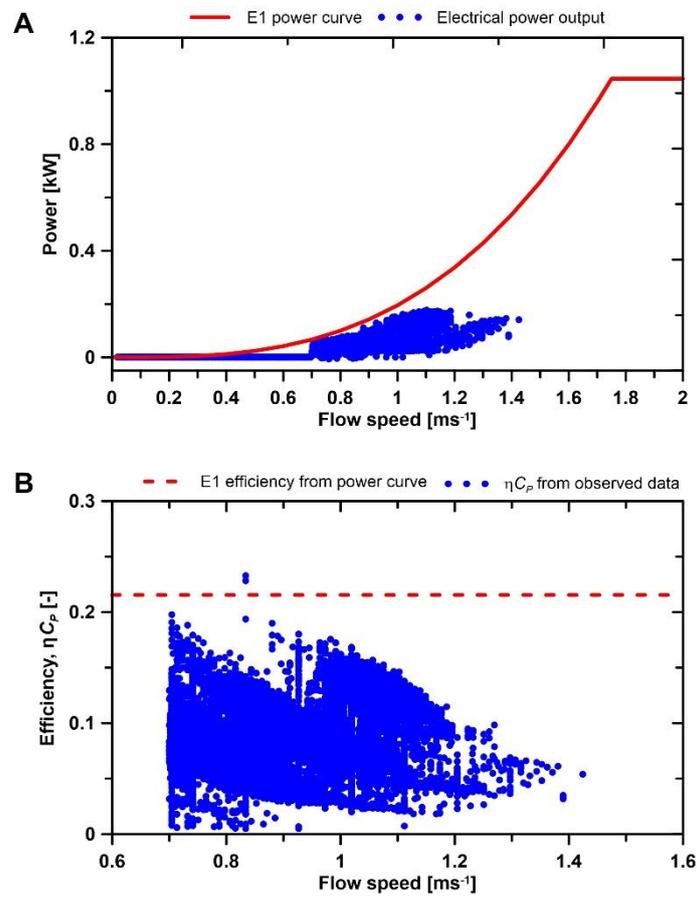

**Figure 15.** (A) Comparison between E1' electrical power curve and the observed electrical power outputs; (B) Observed efficiencies, $\eta C_P$, of E1 at various flow speeds.

**FIGURE 16**

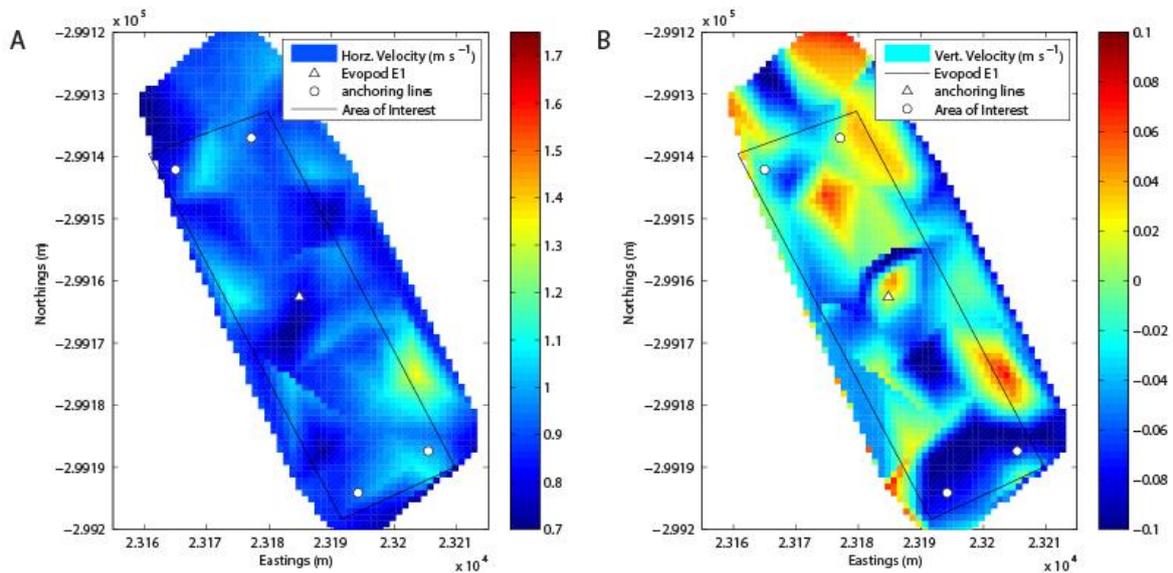

**Figure 16.** Example of the 2D wake profiles (e.g. 0.75-2.25 m from surface) measured with the Sontek 1 MHz during peak flood: (A) A snapshot of the current horizontal velocities at the deployment area (black rectangle) where it can be observed a complex unsteady flow field; and (B) vertical flow velocities showing an increase of the turbulence at the expected wake location. E1 position is marked with a white cross and the four buoys delimiting the area are presented with ta white dot. Note that the colour bar has different scale in the two plots.

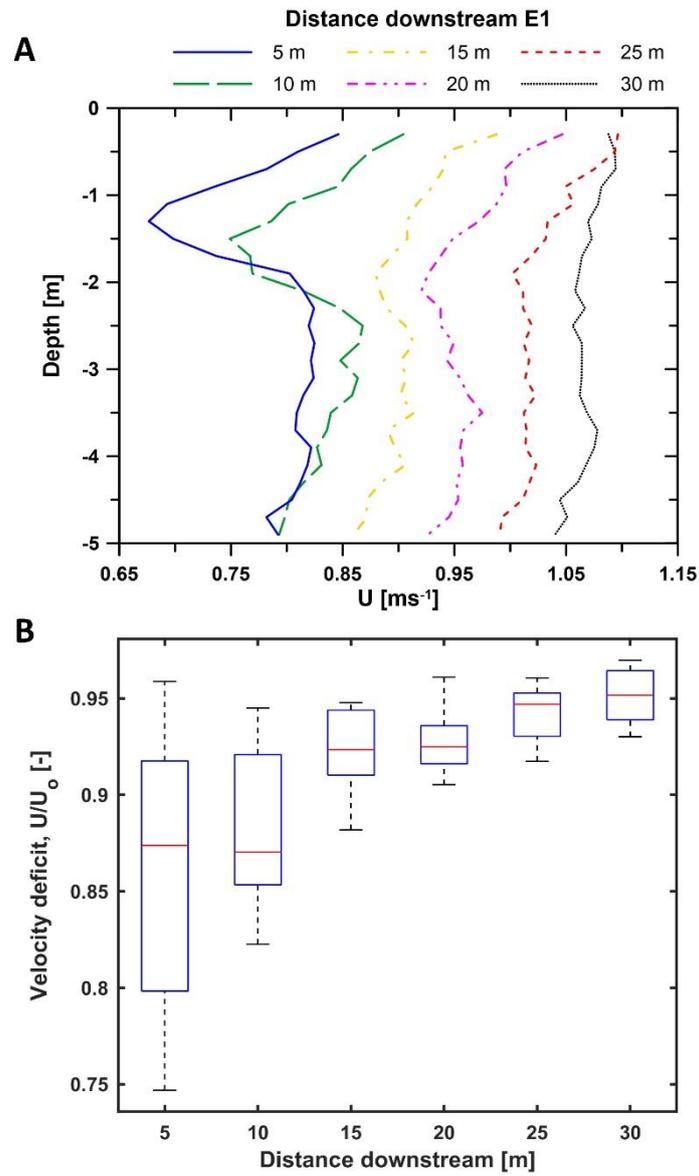

**Figure 17.** (A) Example of the static wake profiles measured with the Nortek AS Signature 1 MHz at different E1' downstream distances; (B) Box plot of wake centreline velocity deficits ($U/U_o$) at E1' rotor height.

# TABLE 1

**Table 1.** Evopod key parameters (adapted from Mackie [6])

|  | **Full scale** (Pentland Firth) | **1:10th scale** (Stranford Narrow IRL / Ria Formosa PT) | **1:40th scale** (Newcastle University test tank, UK) |
|---|---|---|---|
| Length overall (m) | 21.5 | 2.15 | 0.538 |
| Breadth across struts (m) | 13.7 | 1.37 | 0.343 |
| Displacement (t) | 375,000 | 375 | 5.86 |
| Turbine diameter (m) | 15 | 1.5 | 0.375 |
| Rated output (kW) | 1800 | 0.57 | 0.004 |
| Rated flow speed (ms$^{-1}$) | 4.0 | 1.26 | 0.63 |
| Average operating sea state | $H_s$ = 3 m<br>$T_z$ = 8 s | $H_s$ = 0.3 m<br>$T_z$ = 2.5 s | $H_s$ = 0.0075 m<br>$T_z$ = 1.26 s |
| Survival sea state | $H_s$ = 14 m<br>$T_z$ = 14 s | $H_s$ = 1.4 m<br>$T_z$ = 4.43 s | $H_s$ = 0.35 m<br>$T_z$ = 2.21 s |

# TABLE 2

**Table 2.** Tidal stream, wind and wave characteristics used in mooring design. Wave and wind data used on the computations were obtained from the wave buoy offshore Faro-Olhão Inlet and the meteorological weather station of Faro International Airport, respectively.

| | |
|---|---|
| Predicted spring tide peak flow | 1.5 ms$^{-1}$ |
| Percentage time flow exceeds 0.7 ms$^{-1}$ | 20 % |
| Percentage time flow exceeds 1.75 ms$^{-1}$ | 0 % |
| Estimated wind induced surface current | 0.2 ms$^{-1}$ |
| Extreme current speed for mooring design | 1.7 ms$^{-1}$ |
| Wind Direction | NE or NW |
| Wind Speed | 35 kmhr$^{-1}$ (9.7 ms$^{-1}$ or 18.8 knots) |
| Fetch | 4 km (2.2 nautical miles) |
| Significant wave height $H_s$ | 0.45 m |
| Significant wave period $T_{1/3}$ | 2.6 s |
| Mean zero up-crossing period $T_z$ | 2.4 s |

# TABLE 3

Table 3. SCORE database following the European Marine Energy Centre (EMEC) guidelines.

| Type of data | Date of Survey | Method used | Coverage / Resolution | Data format / Availability |
|---|---|---|---|---|
| Bathymetry | 2011 | Lidar | Ria Formosa / 10 m | Netcdf file / under request |
| Bathymetry | 2015 | Single beam echosounder syncronized with a RTK-DGPS / tide corrected | Faro Channel / | Netcdf file / open source |
| Bathymetry | 07/2016 | Single beam echosounder syncronized with a RTK-DGPS / tide corrected | Deployment area / lines spaced every 10 m and depths collected at each 1 sec | Netcdf file / open source |
| Side scan sonar | 07/2016 | Transects performed with a bed imaging system to characterise the bottom of the deployment area in terms of materials and the texture type. | Deployment area | Mosaic image Netcdf file / open source |
| Bed characterization | 07/2016 – 07/2017 | Van Veen dredge operated from the boat. Samples were sieved and benthos organisms, conserved in 98% alcohol for taxonomic identification and counting. | Deployment area / updrift and downdrift of the E1 point location | Pdf document and an excel data file with identified and quantified organisms typical / frequent in the Ria Formosa, as well sediment properties characterization Netcdf file / open source |
| Habitat characterisation | 07/2016 – 07/2017 | Bottom trawling, visual census and ROV images to capture, identify and quantify fish species, invertebrates, and epithelial or benthic species on mobile substrate | Deployment area / updrift and downdrift of the E1 point location | Pdf document and an excel data file with identified and quantified organisms typical / frequent in the Ria Formosa Netcdf file / open source |
| Tidal currents | 03/11/2016 - 17/11/2016 | ADP Nortek Signature 1 MHz - bottom mounted on a frame structure, up looking | Deployment area, 8m depth Avg. Interval: 1min Measur. Int.: 5 min Cell size: 0.2 m Start profile: 0.2m End profile: 8m Coordinate System: ENU | For each cell: time (UTC); ENU velocities; standard deviation in the three directions; signal to noise ratio (SNR) for the three directions; temperature; pressure. Netcdf file / open source |

**Table 3 (cont).** SCORE database following the European Marine Energy Centre (EMEC) guidelines

| Type of data | Date of Survey | Method used | Coverage / Resolution | Data format / Availability |
|---|---|---|---|---|
| Tidal currents | | ADP Sontek 1.5kHz Static survey, down looking | Deployment point, 8m depth Full tidal cycle Avg. Interval: 5 sec Measur. Int.: 5 sec Cell size: 0.5 m Start profile: 0.7 m End profile: 8 m Coordinate System: ENU | For each cell: time (UTC); ENU velocities; standard deviation in the three directions; signal to noise ratio (SNR) for the three directions; temperature; pressure. Netcdf file / open source |
| Acoustic measurements | 19/01/2017 – 14/02/2017 | DigitalHyd SR-1 | Deployment area, 11m depth Sampling rate: 52734 sps Amplitude resol.: 24 bits Avg. Interval: 90s Measur. Int: 10min | Time-series of sound pressure levels (dB) and frequency (kHz) Netcdf file / open source |
| Wake measurements | ?/11/2017 | ADP Nortek Signature 1 MHz Static E1 centreline profiles at: 5 m up-stream, and 5 m, 10 m; 15 m, 20 m, 25 m, and 30 m down-stream. down looking | Boat operated Avg. Interval: 5 sec Measur. Int.: 5 sec Cell size: 0.2 m Start profile: 0.2m End profile: 8m Coordinate System: ENU | For each cell: time (UTC); ENU velocities; standard deviation in the three directions; signal to noise ratio (SNR) for the three directions; temperature; pressure. Netcdf file / open source |
| Wake measurements | | ADP Sontek 1.5kHz Transect survey, E1 transversal profiles 5m spaced within the deployment area. down looking | Boat operated Avg. Interval: 5 sec Measur. Int.: 5 sec Cell size: 0.5 m Start profile: 0.7 m End profile: 8 m Coordinate System: ENU | For each cell: time (UTC); ENU velocities; standard deviation in the three directions; signal to noise ratio (SNR) for the three directions; temperature; pressure. Netcdf file / open source |
| Turbine performance data | 08/06/2017 – 21/11/2017 | Evopod E1 data collection | Deployment point, 8m depth Logging values every 10s | Shaft speed (RPM), load cells (kN), generate voltage (volts), generate amperage (amps), input velocity (ms$^{-1}$), electrical output (W), raw power (W) Netcdf file / open source |

**Table 4.** Pairwise Multiple Comparison Procedures (Tukey Test) of the percent-cover of the arborescent bryozoan *Bugula neritina* observed in the study areas in the four surveyed periods. T-3: 3 days before to the deployment; T+ 8, T+15 and T+63: 8, 15 and 63 days after the turbine had been installed.

|  | Control |  | Turbine |  |
|---|---|---|---|---|
| **Time** | **Diff of Ranks** | **q** | **Diff of Ranks** | **q** |
| **T+63, T-3** | 12967 | 20.882* | 14120 | 17.063* |
| **T+63, T+8** | 14988 | 18.112* | 11707.5 | 14.148* |
| **T+63, T+15** | 6889 | 16.624* | 7052.5 | 8.522* |
| **T+15, T-3** | 6078 | 14.667* | 7067.5 | 8.541* |
| **T+15, T+8** | 8099 | 13.043* | 4655 | 5.625* |
| **T+8, T-3** | 2021 | 4.877 | 2412.5 | 2.915 |

\* $P<0.05$

# TABLE 5

**Table 5.** Issues, problems, consequences and actions during E1 deployment and operational period

| Date | Issue | Problem | Consequence | Action |
|---|---|---|---|---|
| 07/06/2017 | Mooring tension load cells | False tension readings were recorded at three of four load cells before deployment | Loss of drag data. Three load cells need a total rebuild with new electronics | Boat was already commissioned and SCORE team decided to deploy E1 anyway and removed it one month later to repair load cells |
| 08/06/2017 | Deployment | n/a | n/a | Successful |
| 08/06/2017 | GSM Modem | Communication failure | Impossible to download data remotely | Plan extra visits to the E1 to download data via USB cable |
| 14/06/2017 | Battery | Battery failure | Navigation light failed due to lower voltage, Battery failed to charge the logger and E1 stopped to record data | New disposal navigation light was added. New batteries and new charge controller was ordered. |
| 18/06/2017 | E1 underwater | E1 keel while rotating caught the SE mooring line that was over tensioned | Rotor continued to rotate causing a sink force that pulled E1 underwater; water penetrated on the solar panel connectors that were not proper sealed | The diver removed the over tension of the mooring line on the keel. E1 had to be removed from site and towed back to shore for maintenance. |
| 26/06/2017 | Recovered | n/a | n/a | Successful |
| 13/07/2017 | Deployment | Three load cells need a total rebuild with new electronics | Loss of drag data. Three load cells need a total rebuild with new electronics. Sent to factory for repairing. | The solar connectors were fixed. Extra mooring chain and clump weights added to avoid any E1 rotation problem; Place the undamaged load cell at the two North moorings to get tension measurements at the stronger ebb current |
| 22/07/2017 | Logger | The logger was set up in overwrite oldest readings mode | Sampling at 1Hz, Squirrel Logger has enough memory for 28 days. After that starts to overwrite its stored | Planned data retrieval every 20 days |
| 03/08/2017 | Logger | At neap tides not enough flow for the turbine to generate and feed voltage to the logger | Loss of data. Battery voltage fell below 5.5V at which point the logger disarmed and shut down. | Plan the recovery to fit additional solar panels; add a top box at the deck with 2 new lead acid 12V / 5Ah batteries connected to the inside ones |
| 22/08/2017 | Recovered | n/a | n/a | Successful |
|  | Compass | Compass failure | No compass data | Refit compass failed |
|  | Mooring tension load cells | Pot new load cells | Connect up new load cells and test with hang off weights | Two new cells added to be placed on N and S moorings, respectively |
|  | Battery | Not enough power on neap tides to charge the batteries | Fit battery box to top deck and connect into existing wiring; add extra solar panels | More battery power and capacity to charge with the additional solar panels |
| 22/09/2017 | Deployment | n/a | n/a | Successful |
|  | Battery | Not enough power on neap | Intermittent data collection. Gaps on the time-series | Increase extra visits to the site to charge and replace batteries and maintain |
| | | tides to charge the batteries | | |
| 21/11/2017 | Recovered | n/a | n/a | Successful |